# Topological protection in nonlinear optical dynamics with parity-time symmetry

Running Title: Topological dynamics with parity-time symmetry

Sunkyu Yu*, Xianji Piao, and Namkyoo Park*

*Photonic Systems Laboratory, Department of Electrical and Computer Engineering, Seoul National University, Seoul 08826, Korea*

*E-mail address:*

*Sunkyu Yu: skyu.photon@gmail.com*

*Xianji Piao: xjpiao227@gmail.com*

*Namkyoo Park: nkpark@snu.ac.kr*

*The full contact details of the corresponding author:*

Tel: +82-2-880-1820

Fax: +82-2-885-5284

## Abstract

Topological phases exhibit properties that are conserved for continuous deformations, as demonstrated in topological protections in condensed-matter physics and electromagnetic waves. Despite its ubiquitous nature and recent extensions to synthetic dimensions, non-Hermitian Hamiltonians, and nonlinear dynamics, topological protection has generally been described in spatial lattices with the Chern number in the Brillouin zone, focusing on the realization of backscattering-free wave transport. Here, we investigate a different class of topological protection in parity-time-symmetric nonlinear optical dynamics, exploiting the topological invariance of optical state trajectories. For coupled nonlinear photonic systems composed of gain and loss atoms, we classify the topology of equilibria separately for unbroken and broken parity-time symmetry. Utilizing the immunity of topological phases against temporal perturbations, we develop noise-immune laser modulation and rectification with a parasitic nonlinear resonator based on oscillation quenching mechanisms that are protected by parity-time symmetry. The connection between topological photonics and parity-time symmetry through nonlinear dynamics provides a powerful platform for noise-immune signal processing.

## Introduction

The topological degree of freedom in band theory has provided a new phase of matter, such as quantized bulk conductance and topologically protected edge states[1]. The similarity between the Schrödinger equation and Maxwell's wave equations has also stimulated the birth of topological photonics[2, 3]. One of the major goals in this field is to realize a photonic analogy of topological phenomena in condensed-matter physics, which enables backscattering-free light propagation[4, 5]. Topological photonics has recently been extended to synthetic dimensions[6], non-Hermitian photonics[7], and optical nonlinearities[8, 9]. All of these efforts share a common definition of topology: the topological nature of optical wavefunctions in the dispersion band[1, 2].

However, given the ubiquitous property of topology[10], we need to explore other classes of topological invariants in photonics, which are defined for optical quantities or phenomena rather than band structures. An important example is the topology of an isofrequency surface and its phase transition[11, 12] as the optical equivalent of the Lifshitz transition for a Fermi surface[13]. This type of topology enabled the discovery of hyperbolic materials that have provided new design freedom in metamaterials[14]. When considering recent interest in optical dynamics using nonlinearities[15] or time-varying media[16], we also expect the utilization of a certain type of topological invariants in optical dynamics, which will offer further design freedom for wave devices in the temporal domain.

It is well known that the topological equivalence between different dynamical systems is defined by the existence of an invertible map between the state trajectories of each system, i.e., the homeomorphism of phase portraits[17]. Because the topology of the state trajectory of a dynamical system is closely related to the energy exchange between the system and its environment, topological phases in optical dynamics should have a natural connection with non-Hermitian photonics[18]. So far, numerous studies on nonlinear optical dynamics have been conducted for specific types of optical nonlinearities in non-Hermitian systems. For example, a lasing platform, which is a traditional non-Hermitian system with inherent nonlinearity in amplification, has been extensively studied in terms of nonlinear optical dynamics in order to examine chaos, instability, and synchronization in lasing dynamics[19-21]. The recent developments in parity-time (PT) symmetry and topological physics have established new design freedom in nonlinear optical dynamics: the suppression of time reversals[22], optical isolation[23], amplified Fano resonances[24], single-mode lasing[25-27], and quenching behaviors[28] in nonlinear PT-symmetric systems, and topological zero modes in Su-Schrieffer-Heeger chains[29]. However, despite these various achievements, we emphasize that a common and essential analysis for general dynamical systems—topological classification and protection of state trajectories[17]—are still absent in nonlinear PT-symmetric dynamics.

In this paper, we study nonlinear optical dynamics in PT-symmetric systems in terms of topological protection by employing dynamical system theory[17] and thus achieve noise-immune signal processing using topologically protected phases. From the topological class defined by the topological equivalence of optical state trajectories, we show that the topological classification of nonlinear optical dynamics is determined by PT symmetry[18]. This classification also reveals PT-symmetry-protected oscillation quenching mechanisms: amplitude death (AD)[30] and oscillation death (OD)[31] without the assistance of phase delay lines, in sharp contrast to previous approaches[30, 31]. Using topological immunity against temporal perturbations, two application examples are also presented: noise-immune laser modulation and rectification. The extension of topological protection into non-Hermitian photonics and nonlinear dynamics can be readily implemented with other platforms, such as electric circuits and acoustics.

## Results

**Model definition.** Let us consider a generic photonic system coupled to an external reservoir, such as a photonic molecule consisting of two coupled nonlinear resonators each with the same resonance frequency $\omega_0$ and coupling coefficient $\kappa$. The nonlinearity of each resonator is assumed to be any form of an intensity-dependent gain or loss. The photonic molecule is described by temporal coupled mode theory (TCMT)[32] as

$$\frac{da_1}{dt} = i\omega_0 a_1 + N_1(|a_1|^2)a_1 + i\kappa a_2,$$
$$\frac{da_2}{dt} = i\omega_0 a_2 + N_2(|a_2|^2)a_2 + i\kappa a_1, \tag{1}$$

where $a_m$ and $N_m$ represent the field amplitude and real-valued nonlinearity function of the $m^{\text{th}}$ resonator ($m = 1, 2$), respectively. The abstract forms of $N_{1,2}$ represent universal intensity-dependent nonlinearities (Supplementary Note S1): multiphoton processes[33], saturable responses[23], and their arbitrary combinations, dramatically increasing the possible design freedom. We note that although Eq. (1) and its similar form has been widely studied[19-29], the topological classification of optical state trajectories and their protections have not been considered.

To analyze the evolution of the optical energy, we derive an equation for the intensity with $a_m = I_m^{1/2}\exp(i\varphi_m)$ as[34]

$$\frac{dI_1}{dt} = 2N_1(I_1)I_1 + 2\kappa\sqrt{I_1 I_2}\sin\theta,$$
$$\frac{dI_2}{dt} = 2N_2(I_2)I_2 - 2\kappa\sqrt{I_1 I_2}\sin\theta, \tag{2}$$

where $I_m$ and $\varphi_m$ are real-valued intensity and phase functions, respectively, and $\theta = \varphi_1 - \varphi_2$ is the time-varying phase difference between each resonator field. Although $\varphi_{1,2}(t)$ and $\theta(t)$ are determined by coupled time-derivative equations (Supplementary Note S2), at this stage, we consider the steady-state solution with static $I_{1,2}$ and $\theta(t,I_1,I_2) = \theta_s(I_1,I_2)$ near equilibria[17]. This resonator synchronization around equilibria will be discussed later with Supplementary Note S7.

With the synchronization, we assign $da_m/dt = i\omega a_m$ with static $N_m(I_m)$. The static function $\theta_s(I_1,I_2)$ for all eigenmodes of Eq. (1) then satisfies (see Supplementary Note S3)

$$\sin\theta_s = \begin{cases} \gamma, & \text{for } |\gamma| \leq 1, \\ \text{sgn}(\gamma), & \text{for } |\gamma| > 1, \end{cases} \quad (3)$$

where $\gamma = [N_2(I_2) - N_1(I_1)]/(2\kappa)$, and $\text{sgn}(x)$ is the sign function: $\text{sgn}(x \geq 0) = +1$ and $\text{sgn}(x<0) = -1$. The upper and lower conditions in Eq. (3) correspond to unbroken and broken PT symmetry[18], respectively.

**Stability analysis.** To define the topology of our system, we explore the equilibria in Eq. (2) and examine their stability[17]. The equilibrium $(I_{1E}, I_{2E})$ in the two-dimensional (2D) state space $I_1$-$I_2$ is obtained with $dI_{1,2}/dt = 0$ in Eq. (2), which results in

$$N_1(I_{1E})I_{1E} = -N_2(I_{2E})I_{2E} = -\kappa\sqrt{I_{1E}I_{2E}}\sin\theta_s(I_{1E},I_{2E}), \quad (4)$$

with the static function $\theta_s(I_{1E}, I_{2E})$ at equilibrium. Although we initially assign arbitrary nonlinearities, the first equality in Eq. (4) shows that $N_1(I_{1E})$ and $N_2(I_{2E})$ should have different signs for a nontrivial equilibrium $I_{1E,2E} > 0$, which exhibits the connection between nonlinear optical dynamics and PT symmetry with the gain-loss configuration[18].

The stability of the equilibrium $(I_{1E}, I_{2E})$ is examined by the first Lyapunov criterion[17] in which the eigenvalues of the Jacobian matrix of Eq. (2) are used (Supplementary Note S4 for the Jacobian matrix $A$). We note that Eq. (3) provides a separate analysis of each phase of PT symmetry. For the case of unbroken PT symmetry ($|\gamma| \leq 1$, Supplementary Note S5), the system is in the homogeneous steady state (HSS)[30] at equilibrium as $I_{1E} = I_{2E} = I_{HE}$, where $I_{HE}$ is obtained from $N_1(I_{HE}) = -N_2(I_{HE})$. This equilibrium reduces the system dimensionality from 2D to 1D due to the degeneracy of $A$, resulting in a single Jacobian eigenvalue $\lambda_{HE} = [N_1'(I_{HE}) + N_1'(I_{HE})]I_{HE}$, where $N_m' = dN_m/dI_m$. In contrast, for broken PT symmetry ($|\gamma| > 1$, Supplementary Note S6), the equilibrium is determined by the relation $N_1(I_{1E})N_2(I_{2E}) = -\kappa^2$ with the intensity ratio $I_{2E}/I_{1E} = -N_1(I_{1E})/N_2(I_{2E})$, forming an inhomogeneous steady state (IHSS)[31] with $I_{1E} \neq I_{2E}$. The eigenvalues of $A$ are then achieved as $\lambda_{\pm E} = [N_1(I_{1E}) + N_2(I_{2E})]/2 + N_1'(I_{1E})I_{1E} + N_2'(I_{2E})I_{2E} \pm \rho^{1/2}/2$ (Supplementary Note S6 for $\rho$). With Eq. (3) and the equilibrium, the static phase difference condition $\theta(t,I_1,I_2) = \theta_s(I_1,I_2)$ near equilibria is proved in all phases of PT symmetry regardless of the value of $|\gamma|$ (Supplementary Note S7).

**Topological classification.** As an example, we investigate the simple functions of $N_1(I_1) = \eta_{11}I_1 + \eta_{10}$ and $N_2(I_2) = \eta_{20}$, which gives analytical solutions (Supplementary Note S8). In practical systems, $\eta_{11}$ describes two-photon absorption (TPA) or emission (TPE), and $\eta_{10,20}$ represents linear gain or loss. We also assume $\kappa \geq 0$ and a fixed value of $\eta_{11}$. We classify the topological phases of the nonlinear photonic molecule by the topological equivalence of the dynamical trajectories in the 2D state space $I_1$-$I_2$ (i.e., the homeomorphism of phase portraits[17]). According to the Grobman-Hartman theorem[35], the phase portraits of the system near a hyperbolic equilibrium, which does not have Jacobian eigenvalues on the imaginary axis, are locally topologically equivalent to those of its linearized system. Because the topology of this linearized system is quantized by the pair $(n_+,n_-)$[17], where $n_\pm$ denote the numbers of Jacobian eigenvalues with positive and negative real parts, we use $(n_+,n_-)$ as the quantized "topological charge of dynamical systems", analogous to the genus number[10] or Chern number[1-3]. For general 2D systems near hyperbolic equilibria, three topological phases exist according to the phase portrait (Fig. 1a), with their topologies characterized by $(n_+,n_-)$ (Fig. 1b): the S(0,2) stable phase, D(1,1) saddle phase, and U(2,0) unstable phase[17], analogous to negative, zero, and positive electric charges. Although the S and U phases are divided into node ($S^N$,$U^N$) and focus ($S^F$,$U^F$) phases according to their detailed trajectories, the phases with the same ($n_+$,$n_-$) are topologically equivalent[17].

Figure 1c and d shows the topological classification in the parameter space $\eta_{10}$-$\eta_{20}$ for the example of the TPA ($\eta_{11} < 0$) nonlinearity. The Jacobian eigenvalues are obtained for unbroken (Fig. 1c) and broken (Fig. 1d) PT symmetry, except for forbidden regions (gray color, $I_{1E,2E} < 0$, Supplementary Note S9). For unbroken PT symmetry with $|\gamma| \leq 1$ at equilibria, a single topological phase exists: the S(0,1) stable phase. In contrast, for broken PT symmetry with $|\gamma| > 1$ at equilibria, three topological phases of 2D systems exist: the S(0,2), U(2,0), and D(1,1) phases (Supplementary Note S10). Although the Andronov-Hopf (AH) bifurcation occurs between the S and U phases (red dashed line in Fig. 1d), the saddle-node bifurcation does not occur in this example because of the forbidden region (gray color). For the complete realization of the topological phases in Fig. 1a and b, the use of the TPE ($\eta_{11} > 0$) nonlinearity is also required (Supplementary Note S11).

**Topological protection against optical randomness.** Similar to topological protections of the dispersion band[2, 5, 6], the phase portraits of optical states ($I_1$,$I_2$) in our nonlinear system in which the topology is characterized as ($n_+$,$n_-$) are also robust against temporal perturbations. We verify the topologically protected dynamics by testing the robustness of the phase portraits to random light incidences and system perturbations. Of the various topological phases in Fig. 1, we focus on two "stable" phases for practical applications: the S(0,1) and S(0,2) phases, each with unbroken and broken PT symmetry (Supplementary Note S12 for other topological phases).

Figure 2 shows the calculated state trajectories ($I_1$,$I_2$) with random initial conditions around the equilibrium ($I_{1E}$,$I_{2E}$). The initial fields are completely random in their phases and amplitudes. Although the detailed trajectories of the S(0,1) phase and S(0,2) phases are different, the phase portraits of both topological phases converge to their equilibrium ($I_{1E}$,$I_{2E}$) (red circles) regardless of the initial conditions. Furthermore, the stabilization of each topological phase S(0,1) and S(0,2) involves a different type of oscillation quenching phenomenon protected by PT symmetry: amplitude death (AD) in the S(0,1) phase with the HSS ($I_{1E} = I_{2E} = I_{HE}$)[30] in unbroken PT symmetry, and oscillation death (OD) in the S(0,2) phase with the IHSS ($I_{1E} \neq I_{2E}$)[31] in broken PT symmetry. We note that the slower and oscillatory AD convergence (Fig. 2a) and faster and monotonic OD convergence (Fig. 2b) originate from the inherent properties of PT symmetry: the different and identical real parts of the eigenvalues in unbroken and broken PT symmetry, respectively[18].

**Topological protection against system perturbations.** Next, we investigate topological protection against system perturbations. In conventional topological photonics, topological protections of the dispersion band enable backscattering-free wave transport despite a local deformation of the field profiles[4, 5] for spatial perturbations in the system parameters such as a deformation in the lattice constants or rod radii in photonic crystals. As a temporal equivalent, topological protection in optical dynamics leads to a topology of phase portraits that is immune to temporal perturbations in the system parameters, e.g., a linear gain or loss in $\eta_{10}(t)$ and $\eta_{20}(t)$, despite a local deformation of the phase portraits.

Figure 3 presents the phase portraits of the stable phases S(0,1) and S(0,2) with random perturbations in $\eta_{10}(t)$ and $\eta_{20}(t)$ in the temporal domain (black solid lines, for over 30% maximum errors in $\eta_{10}(t)$ and over 70% maximum errors in $\eta_{20}(t)$ in this example). The temporal variations in the system parameters result in a deformation of the local phase portraits of the optical intensities ($I_1$,$I_2$), similar to the deformed field profiles near the spatial defects in backscattering-free transport examples[4, 5]. However, the convergences to the equilibria (red circles) of the stable phases are topologically protected and eventually lead to AD and OD in the S(0,1) and S(0,2) phases, respectively. This result suggests that the systematic laser stabilization can be achieved with the coupling of a parasitic nonlinear resonator (here, resonator 2) to a lasing gain resonator (here, resonator 1), which leads to oscillation quenching mechanisms.

**Noise-immune signal processing.** Analogous to the effect of a negative charge on an electric field, the topology of the stable phases leads to the convergence of the optical states ($I_1,I_2$) to the equilibrium ($I_{1E},I_{2E}$). This topologically protected convergence against random light incidences and system perturbations allows equilibrium-based, noise-immune signal processing in the temporal domain, such as noise-immune laser modulation (Fig. 4) and rectification (Fig. 5).

Firstly, we show noise-immune laser modulation by exploiting the S(0,2) phase that produces OD. The OD modulation is obtained by controlling a linear gain or loss parameter $\eta_{20}$ (Fig. 4a and yellow dashed arrows in Fig. 1c and d), which can be achieved with optical or electrical pumping to quantum dot films[36] or graphene layers[37]. The variation in $\eta_{20}$ leads to a gradual variation in ($I_{1E},I_{2E}$) (Fig. S3b and c in Supplementary Note S9), which allows noise-immune modulation of the laser outputs $|S_{1-}|^2$ and $|S_{2-}|^2$, each determined by $I_{1E}(\eta_{20}(t))$ and $I_{2E}(\eta_{20}(t))$. Figure 4b-g presents the results of the modulation. When we set the spectral noise component in $\eta_{20}(t)$ (Fig. 4b-d, Note S15 for details), from the topologically protected convergence to equilibrium and the $\eta_{20}$-dependent gradual variation in ($I_{1E},I_{2E}$), evident suppression of the noise component in $\eta_{20}(t)$ is observed in Fig. 4f and g for both $|S_{1-}|^2$ and $|S_{2-}|^2$. We note that the noise suppression in the output is a temporal equivalent of the suppression of the defect-induced local field perturbation in conventional topological structures[2-5].

We now show another example with a higher-level functionality using a dynamical transition between AD and OD: a noise-immune half-wave rectifier utilizing the different nature of each oscillation quenching mechanism. The platform for this application is shown in Fig. 5a. The coupled-resonator laser consists of a nonlinear photonic molecule in which each resonator is coupled to a waveguide with a lifetime $\tau_W$. The nonlinearity functions are thus transformed into $N_1(I_1) = \eta_{11}I_1 + \eta_{10} - 1/\tau_W$ and $N_2(I_2) = \eta_{20} - 1/\tau_W$ due to radiation loss (Materials and methods for the TCMT model). The input signal is set to be the control of a linear gain or loss parameter $\eta_{20}$, and the output signal is defined as $|S_{1-}|^2 - |S_{2-}|^2$, the difference between the powers of the outgoing waves through waveguides 1 and 2.

The input signal $\eta_{20}(t)$ changes the topological phases between two regimes (black dashed arrows in Fig. 1c and d): the AD regime (orange parts) protected by unbroken PT symmetry for the S(0,1) phase and the OD regime (blue parts) protected by broken PT symmetry for the S(0,2) phase. AD and OD then lead to distinct output signals: $|S_{1-}|^2 - |S_{2-}|^2 = 0$ for all regimes of AD, and the continuous change in $(I_{1E}, I_{2E})$ in the OD regime results in a modulation of $|S_{1-}|^2 - |S_{2-}|^2$. Such a dynamical transition between the digital AD operation and the analog OD operation, which also accompanies the switching between 1D and 2D dynamics, constitutes the rectification of $\eta_{20}(t)$. We note that the half-wave rectification in Fig. 5a is accompanied with noise immunity according to the topologically protected convergence to the equilibrium $(I_{1E}, I_{2E})$. For the spectral noise component in $\eta_{20}(t)$ (Fig. 5b-d, Supplementary Note S14), the operation of this application is presented in Fig. 5b-g for different noise levels. As expected for faster OD convergence (Fig. 2b), the analog OD modulation exhibits superior noise suppression.

## Discussion

In conclusion, we theoretically studied the topological nature of nonlinear optical dynamics with PT symmetry, which is manifested by the topological invariance of the trajectory in the optical state space. For generic intensity-dependent nonlinearities in coupled photonic systems, we revealed the crucial link between the topological phases of nonlinear optical dynamics and PT symmetry. Along with the topological protection of each phase in 1D and 2D dynamics, we demonstrated two representative oscillation quenching mechanisms, AD and OD, which are protected by unbroken and broken PT symmetry, respectively. We also showed that stable topological phases and the transition between them allow noise-immune signal processing achieved by AD and OD protected by PT symmetry, such as the combination of digital AD and analog OD operations for rectification.

In terms of the extension of the topological charge into general dynamical systems in an analogy to electrical charges, the achieved noise immunity against temporal perturbations is the dynamical equivalent of backscattering-free propagation. This result also suggests a new systematic laser stabilization methodology: the coupling of a parasitic nonlinear resonator to a lasing gain resonator, which will lead to oscillation quenching mechanisms. With the newly found PT-symmetry-protected oscillation quenching mechanisms, we expect that our platform will have significant implications across dynamical optical devices, such as a noise-immune optical memory and electro-optical logic gates.

## Materials and methods

### Temporal coupled mode theory model for a laser platform

The TCMT formulation for the platform of Figs. 4 and 5 in the main text is[32]

$$\frac{da_1}{dt} = i\omega_0 a_1 + \left[N_1\left(|a_1|^2\right) - \frac{1}{\tau_\text{W}}\right]a_1 + i\kappa a_2 + \sqrt{\frac{2}{\tau_\text{W}}} S_{1+},$$

$$\frac{da_2}{dt} = i\omega_0 a_2 + \left[N_2\left(|a_2|^2\right) - \frac{1}{\tau_\text{W}}\right]a_2 + i\kappa a_1 + \sqrt{\frac{2}{\tau_\text{W}}} S_{2+}, \quad (5)$$

where $S_{1+,2+}$ denote the incident waves through each waveguide. The emitted waves $S_{1-,2-}$ from the nonlinear photonic molecule are obtained as

$$S_{1-} = -S_{1+} + \sqrt{\frac{2}{\tau_\text{W}}} a_1,$$

$$S_{2-} = -S_{2+} + \sqrt{\frac{2}{\tau_\text{W}}} a_2. \quad (6)$$

The incident waves $S_{1+,2+}$ are used for the initial excitation of the nonlinear photonic molecule. We thus set $S_{1+,2+} \sim 0$ after a sufficient time from the excitation. Equation (5) then takes the form of Eq. (1) in the main text with the transformed nonlinearity functions $N_1(I_1) = \eta_{11} I_1 + \eta_{10} - 1/\tau_\text{W}$ and $N_2(I_2) = \eta_{20} - 1/\tau_\text{W}$. Accordingly, the topological phases in Fig. 1 in the main text are achieved by assigning an additional linear gain of $1/\tau_\text{W}$ to each resonator. Additionally, the emitted wave intensities are obtained as $|S_{1-,2-}|^2 = (2/\tau_\text{W})|a_{1,2}|^2$ near the equilibrium with $S_{1+,2+} \sim 0$. Therefore, amplitude death (AD) and OD in each topological phase are successfully manifested in the emission intensities $|S_{1-,2-}|^2$.

## Acknowledgements

We acknowledge financial support from the National Research Foundation of Korea (NRF) through the Global Frontier Program (2014M3A6B3063708), the Basic Science Research Program (2016R1A6A3A04009723), and the Korea Research Fellowship Program (2016H1D3A1938069), all funded by the Korean government.

## Conflict of interests

The authors have no conflicts of interest to declare.

## Contributions

S.Y. conceived the presented idea. S.Y. and X.P. developed the theory and performed the computations. N.P. encouraged S.Y. to investigate nonlinear dynamics for topological photonics while supervising the findings of this work. All authors discussed the results and contributed to the final manuscript.

### Data and materials availability

All data is available in the main text or the supplementary materials.

# Figure legends

**Fig. 1. Topological classification of nonlinear photonic molecules.** (a) Topological phases in general 2D dynamical systems determined by state trajectories. (b) Jacobian eigenvalues of each phase. An orange boundary between the $S^F$ and $U^F$ phases represents the AH bifurcation, while the green boundaries around the D(1,1) phase denote the saddle-node bifurcation. The black dashed line divides node and focus phases. (c,d) PT-symmetry-dependent phase diagrams. (c) $\lambda_{HE}$ for unbroken PT symmetry. (d) $\text{Re}[\lambda_{+E}]$ for broken PT symmetry. The red dashed line denotes the AH bifurcation, and the black dashed lines represent the boundaries between the node and focus phases. The yellow and black dashed arrows in (c,d) represent the transitions for the devices in Figs. 4 and 5, respectively. A schematic of the nonlinear photonic molecules is shown in the third quadrant of (c,d). $\eta_{11}/\kappa = -0.5$.

**Fig. 2. Topologically protected phase portraits.** (a,b) Trajectories of $(I_1,I_2)$ for the (a) S(0,1) phase with $(\eta_{10}/\kappa, \eta_{20}/\kappa) = (1.0, -0.5)$, and (b) S(0,2) phase with $(\eta_{10}/\kappa, \eta_{20}/\kappa) = (2.0, -1.5)$. The initial intensities and phases are determined by $I_1(t=0)/I_{1E} = \{1 + 0.5u[0,1]\cos(u[0,2\pi])\}$, $I_2(t=0)/I_{2E} = \{1 + 0.5u[0,1]\sin(u[0,2\pi])\}$, and $\theta(t=0) = u[0,2\pi]$, where $u[p,q]$ is the uniform random function between $p$ and $q$. $\eta_{11}/\kappa = -0.5$. The trajectories are obtained by solving Eq. (1) using the 6$^{\text{th}}$-order Runge-Kutta method.

**Fig. 3. Topological protections against system perturbations in the temporal domain.** (a,b) Trajectories of $(I_1,I_2)$ for the (a) S(0,1) phase around $(\eta_{10}/\kappa, \eta_{20}/\kappa) = (1.0, -0.5)$ and (b) S(0,2) phase around $(\eta_{10}/\kappa, \eta_{20}/\kappa) = (2.0, -1.5)$. The temporal perturbations in $\eta_{10}(t)$ and $\eta_{20}(t)$ are illustrated as black solid lines (Supplementary Note S13 for exact values of $\eta_{10}$ and $\eta_{20}$). All other parameters are the same as those in Fig. 2.

**Fig. 4. Noise-immune laser modulation.** (a) Schematic of the platform. (b-g) Noise-suppressed laser modulations: (b-d) $\kappa$-normalized modulation signals $\eta_{20}(t)$ with different noise levels and (e-g) output signals $|S_{1-}|^2$ and $|S_{2-}|^2$. The signal in (b) and the yellow lines in (c,d) denote the signal without noise. The S(0,2) phase is maintained for the transition between $(\eta_{10}/\kappa, \eta_{20}/\kappa) = (2.0, -1.1)$ and $(2.0, -2.0)$. $\eta_{11}/\kappa = -0.5$.

**Fig. 5. Noise-immune laser rectification.** (a) Operation principles of the rectifications: AD in the S(0,1) phase and OD in the S(0,2) phase. (b-g) Noise-suppressed laser rectifications: (b-d) $\kappa$-normalized modulation signals $\eta_{20}(t)$ with different noise levels and (e-g) output signals $|S_{1-}|^2 -$

$|S_{2-}|^2$. The signal in (b) and yellow lines in (c,d) denote the signal without noise. $\eta_{11}/\kappa = -0.5$. The initial field is excited once through $S_{1+}$. The TCMT model of Supplementary Note S14 is applied.

# Figures

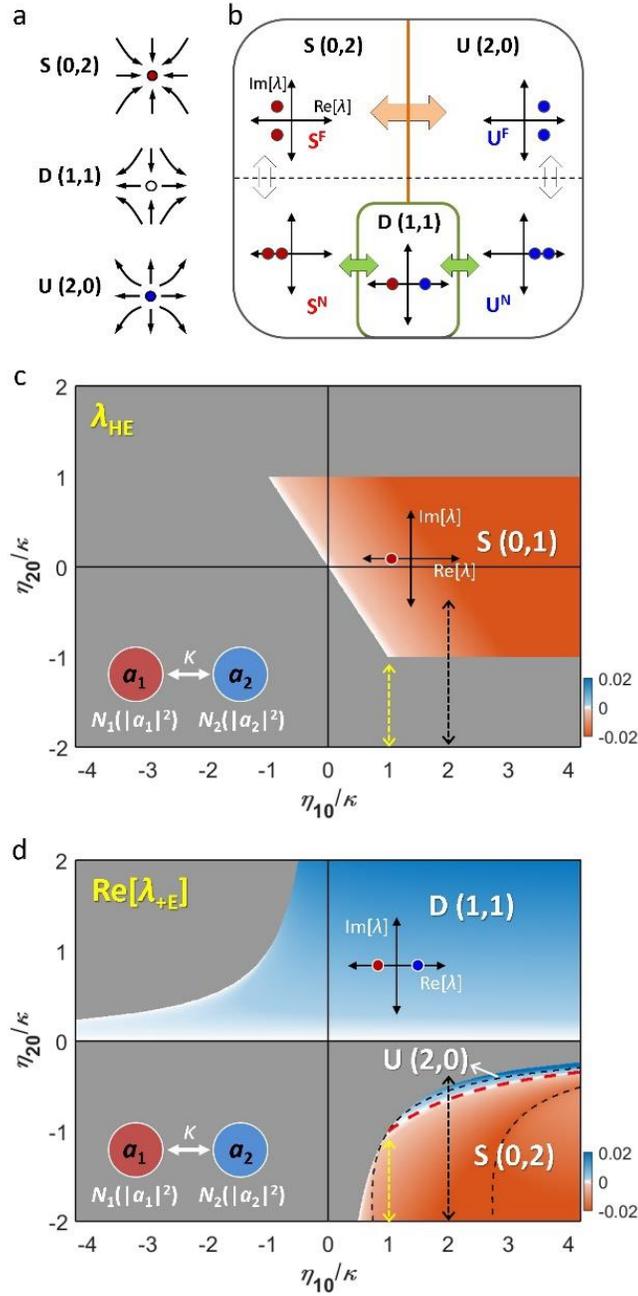

**Figure 1**

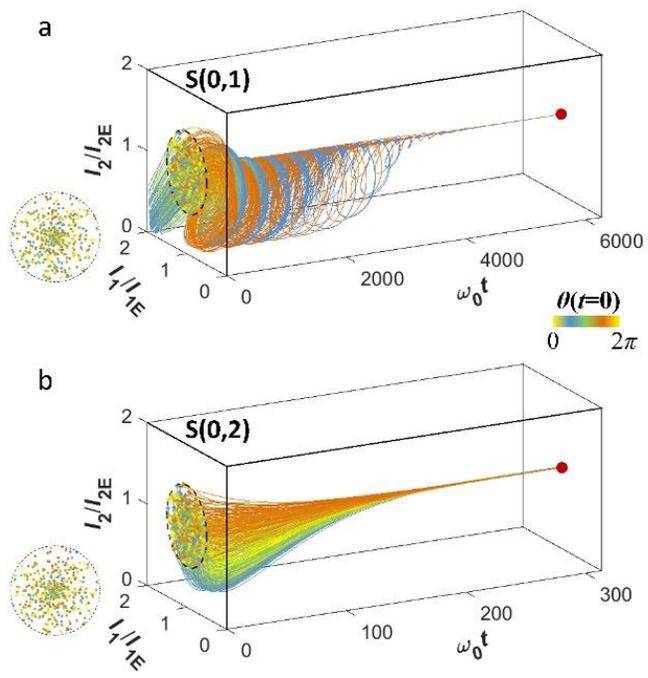

**Figure 2**

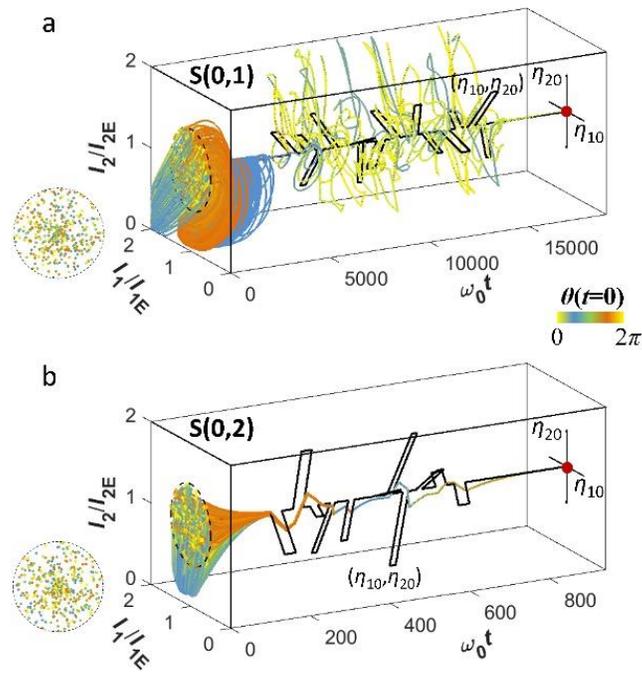

**Figure 3**
24

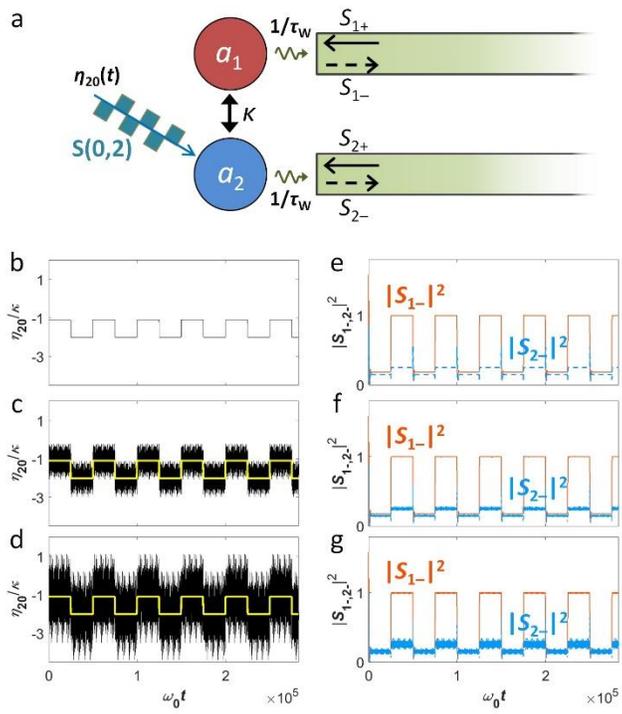

**Figure 4**

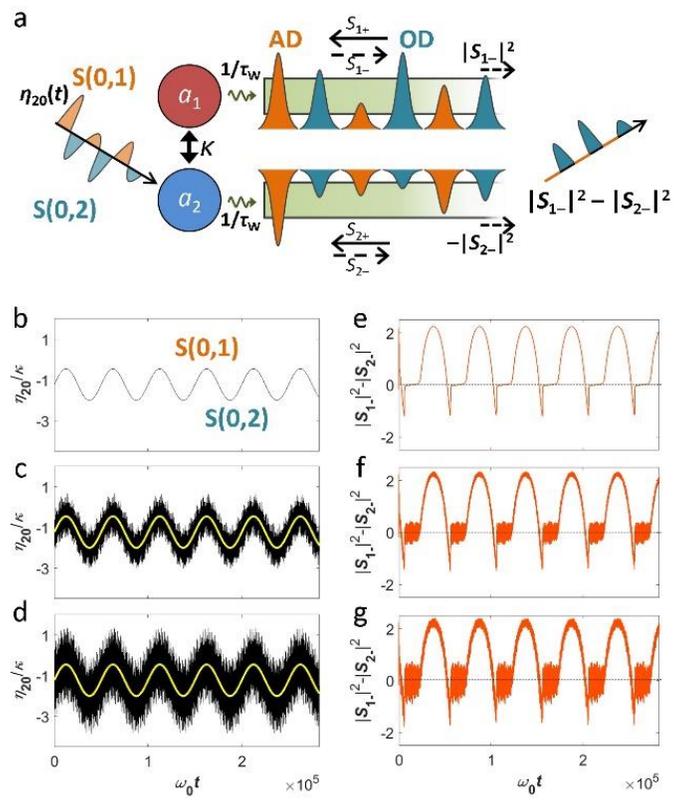

**Figure 5**

**Supplementary Information for "Topological protection in nonlinear optical dynamics with parity-time symmetry"**

Sunkyu Yu*, Xianji Piao, and Namkyoo Park*

Photonic Systems Laboratory, Department of Electrical and Computer Engineering, Seoul National University, Seoul 08826, Korea

*E-mail address for correspondence: nkpark@snu.ac.kr (N.P.); skyu.photon@gmail.com (S.Y.)

**Note S1. Types of intensity-dependent nonlinearity functions**

**Note S2. Phase equations**

**Note S3. Phase difference function**

**Note S4. Jacobian matrix at equilibrium**

**Note S5. Equilibrium and Jacobian matrix of unbroken PT symmetry**

**Note S6. Equilibrium and Jacobian matrix of broken PT symmetry**

**Note S7. Proof of static phase difference**

**Note S8. Analytical solutions of the example system**

**Note S9. Equilibria in the TPA example**

**Note S10. Jacobian eigenvalues of broken PT symmetry in the TPA example**

**Note S11. The TPE example**

**Note S12. Topologically protected dynamics in the saddle and unstable phases**

**Note S13. Temporal system perturbations**

**Note S14. Noise in time-varying loss**

## Note S1. Types of intensity-dependent nonlinearity functions

Intensity-dependent nonlinearities can be classified into saturable responses [1,2] and multiphoton processes [3,4] such as two- or three-photon absorption (TPA [5] or 3PA [6]) and emission (TPE [7] or 3PE [8]). Saturable responses indicate a decrease in a gain or loss coefficient as the light intensity increases, which is observed in the gain saturation of lasers and amplifiers [1,9] and saturable absorption [2,10,11]. These responses are described by the nonlinearity function $N(I) = \eta / [1 + (I/I_s)]$, where $\eta > 0$ for saturable gain, $\eta < 0$ for saturable loss, and $I_s$ denotes the saturation intensity. On the other hand, multiphoton absorption and emission processes exhibit nonlinearity functions with polynomial expressions, as shown in the TPA and TPE with $N(I) = \eta I$ [5,7] and the 3PA and 3PE with $N(I) = \eta I^2$ [6,8], where $\eta > 0$ for emission and $\eta < 0$ for absorption. Figure S1 shows the $N(I)$ functions for different types of intensity-dependent optical nonlinearities.

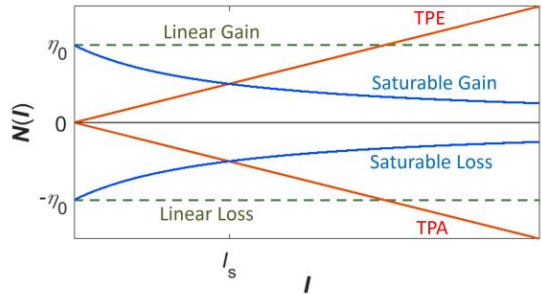

**Fig. S1.** Nonlinearity functions $N(I)$ of intensity-dependent optical nonlinearities: red lines for TPA and TPE with $N(I) = \pm\eta_0 I$ and blue lines for saturable gain and loss with $N(I) = \pm\eta_0 / [1 + (I/I_s)]$. The green dashed lines represent linear gain and loss with $N(I) = \pm\eta_0$. We set $\eta_0 > 0$ for all cases.

**Note S2. Phase equations**

By replacing $a_m$ with $I_m^{1/2}\exp(i\varphi_m)$ in Eq. (1) and employing Eq. (2) in the main text, we achieve

$$\sqrt{I_1}\frac{d\varphi_1}{dt} = \omega_0\sqrt{I_1} - iN_1\sqrt{I_1} + \kappa\sqrt{I_2}e^{-i\theta} + i\frac{N_1 I_1 + \kappa\sqrt{I_1 I_2}\sin\theta}{\sqrt{I_1}},$$
$$\sqrt{I_2}\frac{d\varphi_2}{dt} = \omega_0\sqrt{I_2} - iN_2\sqrt{I_2} + \kappa\sqrt{I_1}e^{+i\theta} + i\frac{N_2 I_2 - \kappa\sqrt{I_1 I_2}\sin\theta}{\sqrt{I_2}}.$$
(S1)

Because the imaginary parts of Eq. (S1) are zero, an equation for the field phase $\varphi_m$ inside each resonator is derived, as

$$\sqrt{I_1}\frac{d\varphi_1}{dt} = \omega_0\sqrt{I_1} + \kappa\sqrt{I_2}\cos\theta,$$
$$\sqrt{I_2}\frac{d\varphi_2}{dt} = \omega_0\sqrt{I_2} + \kappa\sqrt{I_1}\cos\theta.$$
(S2)

The time derivatives of the phase functions $d\varphi_{1,2}(t)/dt$ represent the instantaneous frequencies of the fields in resonators 1 and 2, correspondingly. The static condition of the phase difference $\theta(t) = \varphi_1(t) - \varphi_2(t) = \theta_s$ is found in Eq. (S2), which will be discussed later in Note S7.

## Note S3. Phase difference function

With the resonator synchronization that achieves $da_m/dt = i\omega a_m$ and static $N_m(I_m)$, Eq. (1) in the main text becomes an eigenvalue equation of general PT-symmetric two-level systems [12,13]

$$\begin{bmatrix} \omega_0 - iN_1(I_1) & \kappa \\ \kappa & \omega_0 - iN_2(I_2) \end{bmatrix} \begin{bmatrix} a_1 \\ a_2 \end{bmatrix} = \omega \begin{bmatrix} a_1 \\ a_2 \end{bmatrix}. \tag{S3}$$

From Eq. (S3), $a_1/a_2$ of each eigenmode is obtained as

$$\frac{a_1}{a_2} = i\frac{(N_2 - N_1)}{2\kappa} \pm \sqrt{1 - \left[\frac{(N_2 - N_1)}{2\kappa}\right]^2}, \tag{S4}$$

which is plotted in Fig. S2 with various values of parameter $\gamma = [N_2(I_2) - N_1(I_1)]/(2\kappa)$.

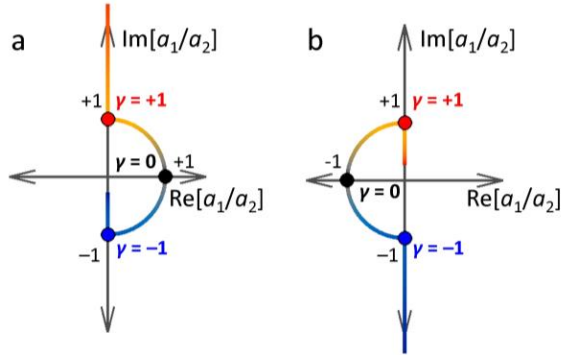

**Fig. S2.** The ratio between each resonator field near equilibria. (a,b) The ratio $a_1/a_2$ of two steady-state eigenmodes with various values of $\gamma$. The blue curve is for $\gamma < 0$, and the red curve is for $\gamma \geq 0$.

Depending on the magnitude of $\gamma$, Eq. (S4) is divided into

$$\frac{a_1}{a_2} = \begin{cases} i\gamma \pm \sqrt{1-\gamma^2} & \text{if } |\gamma| \leq 1, \\ i\left(\gamma \pm \sqrt{\gamma^2 - 1}\right) & \text{if } |\gamma| > 1, \end{cases} \tag{S5}$$

with $|a_1/a_2| = 1$ for $|\gamma| \leq 1$ (unbroken PT symmetry) and $|a_1/a_2| = |\gamma \pm (\gamma^2 - 1)^{1/2}|$ for $|\gamma| > 1$ (broken PT symmetry). From $a_1/a_2 = |a_1/a_2|\exp(i\theta_s)$, we obtain $\sin\theta_s = |a_2/a_1|\text{Im}(a_1/a_2)$, which leads to Eq. (3) in the main text. Although $\theta_s$, the angle in the complex plane in Fig. S2, depends on $N_{1,2}(I_{1,2})$, the values of $\sin\theta_s = \sin(\pi - \theta_s)$ in both eigenmodes are always the same.

**Note S4. Jacobian matrix at equilibrium**

From Eq. (2) of the form $dI_1/dt = F_1(I_1, I_2)$ and $dI_2/dt = F_2(I_1, I_2)$ in the main text, each component of the 2×2 Jacobian matrix $A$ is $A_{i,j} = \partial F_i/\partial I_j$, leading to

$$A = \begin{bmatrix} 2N_1 + 2N_1'I_1 + \kappa\sqrt{\frac{I_2}{I_1}}\sin\theta + 2\kappa\sqrt{I_1 I_2}\frac{\partial \sin\theta}{\partial I_1} & \kappa\sqrt{\frac{I_1}{I_2}}\sin\theta + 2\kappa\sqrt{I_1 I_2}\frac{\partial \sin\theta}{\partial I_2} \\ -\kappa\sqrt{\frac{I_2}{I_1}}\sin\theta - 2\kappa\sqrt{I_1 I_2}\frac{\partial \sin\theta}{\partial I_1} & 2N_2 + 2N_2'I_2 - \kappa\sqrt{\frac{I_1}{I_2}}\sin\theta - 2\kappa\sqrt{I_1 I_2}\frac{\partial \sin\theta}{\partial I_2} \end{bmatrix}. \quad (S6)$$

When the state of the system approaches equilibrium, the time-varying phase difference $\theta(t)$ and the related derivative $\partial \sin\theta(t, I_1, I_2)/\partial I_m$ converge to the static function $\theta_s(I_1, I_2)$ and the derivative $\partial \sin\theta_s(I_1, I_2)/\partial I_m$, respectively. At equilibrium ($I_1 = I_{1E}$ and $I_2 = I_{2E}$), Eq. (4) in the main text derives

$$\begin{aligned}\kappa\sqrt{\frac{I_{2E}}{I_{1E}}}\sin\theta_s(I_{1E}, I_{2E}) &= -N_1(I_{1E}), \\ \kappa\sqrt{\frac{I_{1E}}{I_{2E}}}\sin\theta_s(I_{1E}, I_{2E}) &= N_2(I_{2E}).\end{aligned} \quad (S7)$$

By substituting Eq. (S7) into Eq. (S6) and using the $\theta_s$ function from the equilibrium condition, the Jacobian matrix $A$ at equilibrium ($I_{1E}, I_{2E}$) is obtained, as

$$A = \begin{bmatrix} N_1(I_{1E}) + 2N_1'(I_{1E})I_{1E} + 2\kappa\sqrt{I_{1E}I_{2E}}\frac{\partial \sin\theta_s}{\partial I_1}\bigg|_{I_{1E,2E}} & N_2(I_{2E}) + 2\kappa\sqrt{I_{1E}I_{2E}}\frac{\partial \sin\theta_s}{\partial I_2}\bigg|_{I_{1E,2E}} \\ N_1(I_{1E}) - 2\kappa\sqrt{I_{1E}I_{2E}}\frac{\partial \sin\theta_s}{\partial I_1}\bigg|_{I_{1E,2E}} & N_2(I_{2E}) + 2N_2'(I_{2E})I_{2E} - 2\kappa\sqrt{I_{1E}I_{2E}}\frac{\partial \sin\theta_s}{\partial I_2}\bigg|_{I_{1E,2E}} \end{bmatrix}. \quad (S8)$$

**Note S5. Equilibrium and Jacobian matrix of unbroken PT symmetry**

For unbroken PT symmetry that satisfies $\sin\theta_E(I_{1E}, I_{2E}) = [N_2(I_{2E}) - N_1(I_{1E})]/(2\kappa)$, Eq. (4) in the main text becomes

$$N_1(I_{1E})I_{1E} = -N_2(I_{2E})I_{2E} = -\frac{\sqrt{I_{1E}I_{2E}}}{2}[N_2(I_{2E}) - N_1(I_{1E})]. \tag{S9}$$

The emergence of a nontrivial equilibrium with $I_{1E,2E} \neq 0$ and $N_{1,2} \neq 0$ requires $I_{1E} = I_{2E} = I_{HE}$, which represents the homogeneous steady state (HSS) [14]. The obtained HSS condition simplifies Eq. (S9) to $N_1(I_{HE}) = -N_2(I_{HE})$, which corresponds to balanced nonlinear gain and loss. The intensity value $I_{HE}$ of the equilibrium is then obtained from the specific mathematical forms of $N_1$ and $N_2$, which are determined by the type of optical nonlinearity. The condition of unbroken PT symmetry is also simplified as $|\kappa| \geq |N_1(I_{HE})| = |N_2(I_{HE})|$.

Using the conditions of $I_{1E} = I_{2E} = I_{HE}$, $\partial\sin\theta_E/\partial I_1 = -N_1'/(2\kappa)$, and $\partial\sin\theta_E/\partial I_2 = N_2'/(2\kappa)$, the Jacobian matrix $A$ for unbroken PT symmetry is derived from Eq. (S8) as

$$A = \begin{bmatrix} N_1(I_{HE}) + N_1'(I_{HE})I_{HE} & N_2(I_{HE}) + N_2'(I_{HE})I_{HE} \\ N_1(I_{HE}) + N_1'(I_{HE})I_{HE} & N_2(I_{HE}) + N_2'(I_{HE})I_{HE} \end{bmatrix}. \tag{S10}$$

For nontrivial equilibria ($I_{HE} \neq 0$), Eq. (S10) results in a single eigenvalue $\lambda_{HE} = [N_1'(I_{HE}) + N_1'(I_{HE})]I_{HE}$ due to the degeneracy of the matrix $A$ with the HSS condition $N_1(I_{HE}) = -N_2(I_{HE})$.

**Note S6. Equilibrium and Jacobian matrix of broken PT symmetry**

For broken PT symmetry that satisfies $\sin\theta_E(I_{1E}, I_{2E}) = \text{sgn}([N_2(I_{2E}) - N_1(I_{1E})]/(2\kappa))$, Eq. (4) in the main text becomes

$$N_1(I_{1E})I_{1E} = -N_2(I_{2E})I_{2E} = \begin{cases} +\kappa\sqrt{I_{1E}I_{2E}} & \text{if } \dfrac{N_2(I_{2E}) - N_1(I_{1E})}{2\kappa} < 0 \\ -\kappa\sqrt{I_{1E}I_{2E}} & \text{if } \dfrac{N_2(I_{2E}) - N_1(I_{1E})}{2\kappa} \geq 0 \end{cases}, \quad (S11)$$

It is noted that Eq. (S11) leads to the condition of $N_1(I_{1E})N_2(I_{2E}) = -\kappa^2$. The intensity ratio at the equilibrium $I_{2E}/I_{1E} = -N_1(I_{1E})/N_2(I_{2E})$ does not need to be unity, generally resulting in the inhomogeneous steady state (IHSS) [15]. Notably, the condition of broken PT symmetry is automatically satisfied with $N_1(I_{1E})N_2(I_{2E}) = -\kappa^2$, regardless of the value of $\kappa$.

For broken PT symmetry, $\sin\theta_E$ is constant (Eq. (3) in the main text); thus, $\partial\sin\theta_E/\partial I_1 = \partial\sin\theta_E/\partial I_2 = 0$. The Jacobian matrix $A$ for broken PT symmetry is derived from Eq. (S8) as

$$A = \begin{bmatrix} N_1(I_{1E}) + 2N_1'(I_{1E})I_{1E} & N_2(I_{2E}) \\ N_1(I_{1E}) & N_2(I_{2E}) + 2N_2'(I_{2E})I_{2E} \end{bmatrix}, \quad (S12)$$

which has two eigenvalues $\lambda_{\pm E} = [N_1(I_{1E}) + N_2(I_{2E})]/2 + N_1'(I_{1E})I_{1E} + N_2'(I_{2E})I_{2E} \pm \rho^{1/2}/2$, and $\rho$ is defined as

$$\rho = \left[N_1(I_{1E}) - N_2(I_{2E}) + 2\left(N_1'(I_{1E})I_{1E} - N_2'(I_{2E})I_{2E}\right)\right]^2 + 4N_1(I_{1E})N_2(I_{2E}). \quad (S13)$$

**Note S7. Proof of static phase difference**

In the analysis of the equilibrium and its stability, we employed the static phase difference condition $\theta(t,I_1,I_2) = \theta_s(I_1,I_2)$ for separate analysis of the phases of PT symmetry. This condition requires $d\theta/dt = d(\varphi_1 - \varphi_2)/dt = 0$. From Eq. (S2), we achieve

$$\frac{d(\varphi_1 - \varphi_2)}{dt} = \kappa\left(\sqrt{\frac{I_2}{I_1}} - \sqrt{\frac{I_1}{I_2}}\right)\cos\theta. \tag{S14}$$

Therefore, $d\theta/dt = 0$ for the equilibria in all phases of PT symmetry, due to the conditions of (i) $I_{1E} = I_{2E} = I_{HE}$ for unbroken PT symmetry and (ii) $\cos\theta_E(I_{1E}, I_{2E}) = 0$ from $\sin\theta_E(I_{1E}, I_{2E}) = \pm 1$ for broken PT symmetry.

**Note S8. Analytical solutions of the example system**

For the set of the nonlinearity functions $N_1(I_1) = \eta_{11}I_1 + \eta_{10}$ and $N_2(I_2) = \eta_{20}$, the relation $N_1(I_{HE}) = -N_2(I_{HE})$ for unbroken PT symmetry derives the HSS equilibrium, as $I_{HE} = -(\eta_{10} + \eta_{20})/\eta_{11}$. Due to $N_1'(I_{HE}) = \eta_{11}$ and $N_2'(I_{HE}) = 0$, the Jacobian eigenvalue becomes $\lambda_{HE} = -(\eta_{10} + \eta_{20})$. The necessary condition of unbroken PT symmetry $|\kappa| \geq |N_1(I_{HE})| = |N_2(I_{HE})|$ also becomes $\kappa \geq |\eta_{20}|$ with $\kappa \geq 0$.

The relations of $N_1(I_{1E})N_2(I_{2E}) = -\kappa^2$ and $I_{2E}/I_{1E} = -N_1(I_{1E})/N_2(I_{2E})$ for broken PT symmetry result in the IHSS equilibrium as

$$I_{1E} = -\frac{1}{\eta_{11}}\left(\frac{\kappa^2}{\eta_{20}} + \eta_{10}\right),$$

$$I_{2E} = -\frac{\kappa^2}{\eta_{11}\eta_{20}^2}\left(\frac{\kappa^2}{\eta_{20}} + \eta_{10}\right). \tag{S15}$$

The Jacobian eigenvalues for the equilibrium are then achieved as

$$\lambda_{\pm E} = -\eta_{10} + \frac{\eta_{20}}{2} - \frac{3}{2}\frac{\kappa^2}{\eta_{20}} \pm \frac{1}{2}\sqrt{\left(2\eta_{10} + \eta_{20} + 3\frac{\kappa^2}{\eta_{20}}\right)^2 - 4\kappa^2}. \tag{S16}$$

**Note S9. Equilibria in the TPA example**

Figure S3 shows the equilibria of unbroken (Fig. S3a, $I_{HE}$) and broken (Fig. S3b and c, each for $I_{1E}$ and $I_{2E}$) PT symmetry. In the parameter space $\eta_{10}$-$\eta_{20}$, the nontrivial equilibria exist except for the gray areas, which represent the "forbidden regions" of nontrivial equilibria. First, due to the necessary condition $\kappa \geq |\eta_{20}|$ for unbroken PT symmetry, the equilibrium in Fig. S3a is defined only in the range of $-1 \leq (\eta_{20}/\kappa) \leq 1$, while broken PT symmetry is automatically satisfied with the equilibrium condition $N_1(I_{1E})N_2(I_{2E}) = -\kappa^2$. The other forbidden regions originate from negative intensity values ($I_{HE} < 0$ and $I_{1E,2E} < 0$). The equilibrium in broken PT symmetry satisfies $I_{1E} \neq I_{2E}$ except for $\kappa = |\eta_{20}|$ shown in Eq. (S15), leading to IHSS equilibria.

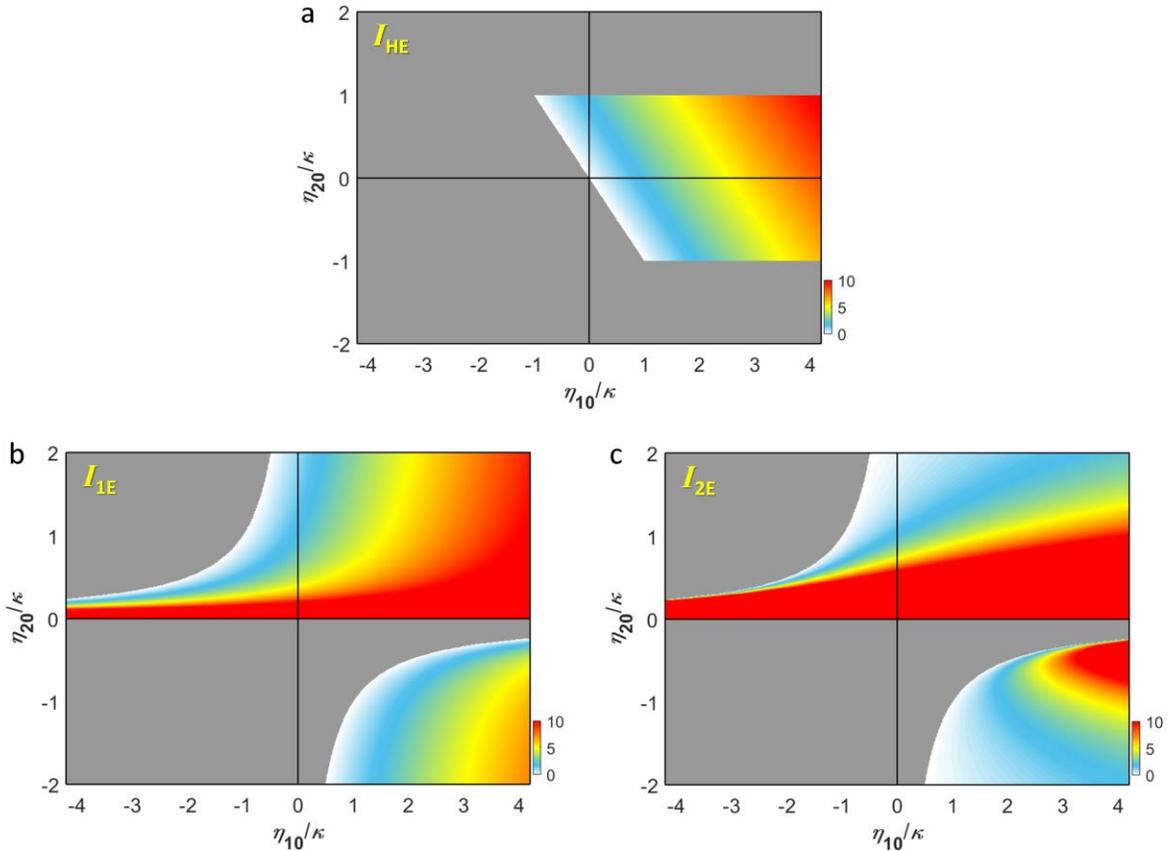

**Fig. S3.** Equilibria in the system parameter space $\eta_{10}$-$\eta_{20}$. (a) $I_{HE}$ for unbroken PT symmetry. (b) $I_{1E}$ and (c) $I_{2E}$ for broken PT symmetry. The gray areas denote the forbidden regions, originating from unbroken PT symmetry and nonnegative intensity values. $\eta_{11}/\kappa = -0.5$ for all cases.

**Note S10. Jacobian eigenvalues of broken PT symmetry in the TPA example**

Figure S4 shows the imaginary part of $\lambda_{+E}$ (Fig. S4a) and complex-valued $\lambda_{-E}$ (Fig. S4b and c) to provide all the information of $\lambda_{\pm E}$ for broken PT symmetry with Fig. 1d in the main text. In the region of $\eta_{20}/\kappa > 0$ except for the forbidden region (gray color), the saddle (D) phase of $(n_+, n_-) = (1,1)$ consists of $\text{Re}[\lambda_{+E}] < 0$ (Fig. 1d in the main text) with $\text{Im}[\lambda_{+E}] = 0$ and $\text{Re}[\lambda_{-E}] > 0$ with $\text{Im}[\lambda_{-E}] = 0$.

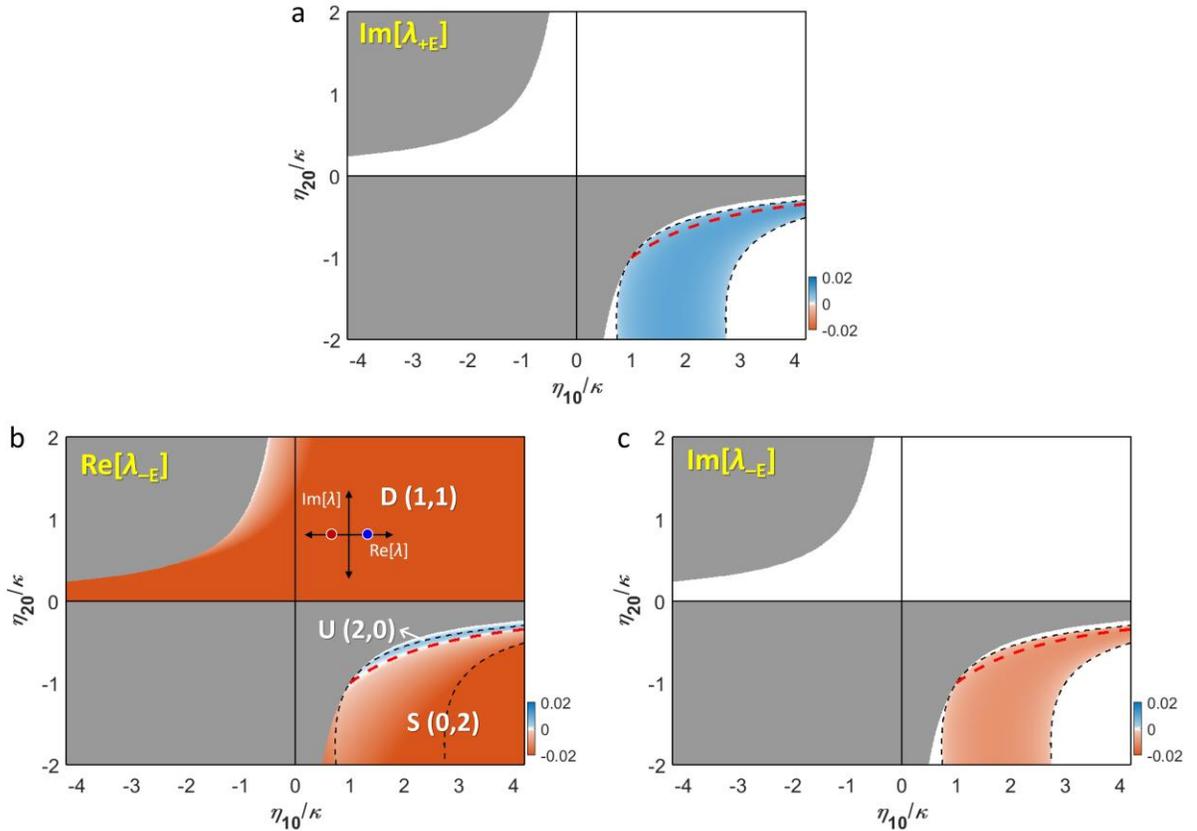

**Fig. S4.** Complex-valued Jacobian eigenvalues. (a) $\text{Im}[\lambda_{+E}]$, (b) $\text{Re}[\lambda_{-E}]$, and (c) $\text{Im}[\lambda_{-E}]$. The red lines denote the AH bifurcation, and the black dashed lines represent the boundaries between the node and focus phases. $\eta_{11}/\kappa = -0.5$ for all cases.

The S and U phases in the region of $\eta_{20}/\kappa < 0$ can be further classified by the types of phase portraits (Fig. S5): node ($S^N$ and $U^N$) and focus ($S^F$ and $U^F$) phases, which are topologically equivalent [16] if the phases have the same $(n_+, n_-)$. While node phases have

Jacobian eigenvalues on the real axis, focus phases have a complex-conjugate pair of eigenvalues. It is noted that the AH bifurcation occurs only between the focus phases $S^F$ and $U^F$.

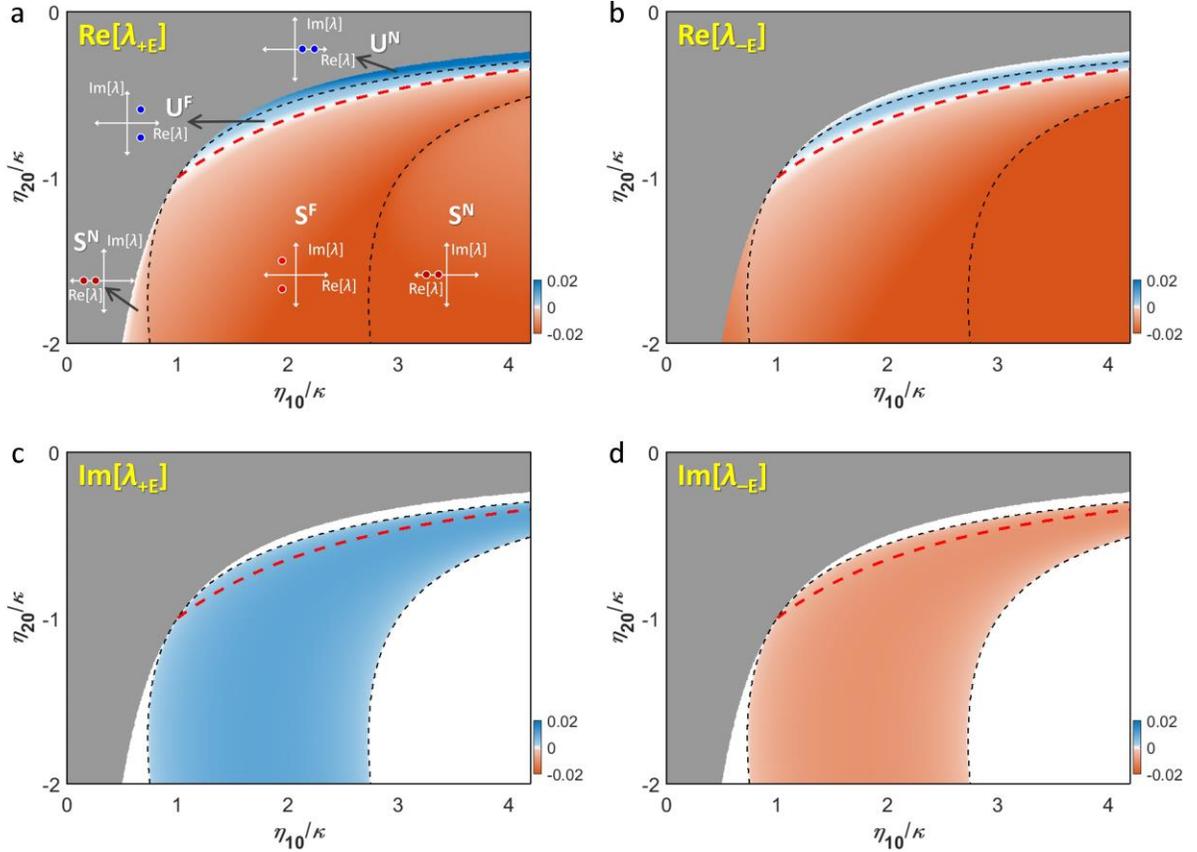

**Fig. S5.** Node-focus classifications in the S and U phases. (a) Re[$\lambda_{+E}$], (b) Re[$\lambda_{-E}$], (c) Im[$\lambda_{+E}$], and (d) Im[$\lambda_{-E}$]. The red lines denote the AH bifurcation, and the black dashed lines represent the boundaries between the node and focus phases. $\eta_{11}/\kappa = -0.5$ for all cases.

## Note S11. The TPE example

We also analyze photonic molecules with the TPE nonlinearity having $\eta_{11} > 0$ (Figs. S6 and S7), which lead to a topological classification that contrasts with the TPA example: a single U(1,0) phase in unbroken PT symmetry and broad U(2,0) and narrow S(0,2) phases in broken PT symmetry. As shown in the analytical solutions $\lambda_{HE} = -(\eta_{10} + \eta_{20})$ and $\lambda_{\pm E}$ in Eq. (S16), the Jacobian eigenvalues in unbroken and broken PT symmetry are independent of $\eta_{11}$. Instead, the values of the equilibria $I_{HE}$ and $I_{1E,2E}$ depend on $\eta_{11}$, which determines the forbidden regions of nontrivial equilibria from negative intensity values ($I_{HE} < 0$ and $I_{1E,2E} < 0$) and thus results in differences between the topological phases of the TPE and TPA photonic molecules.

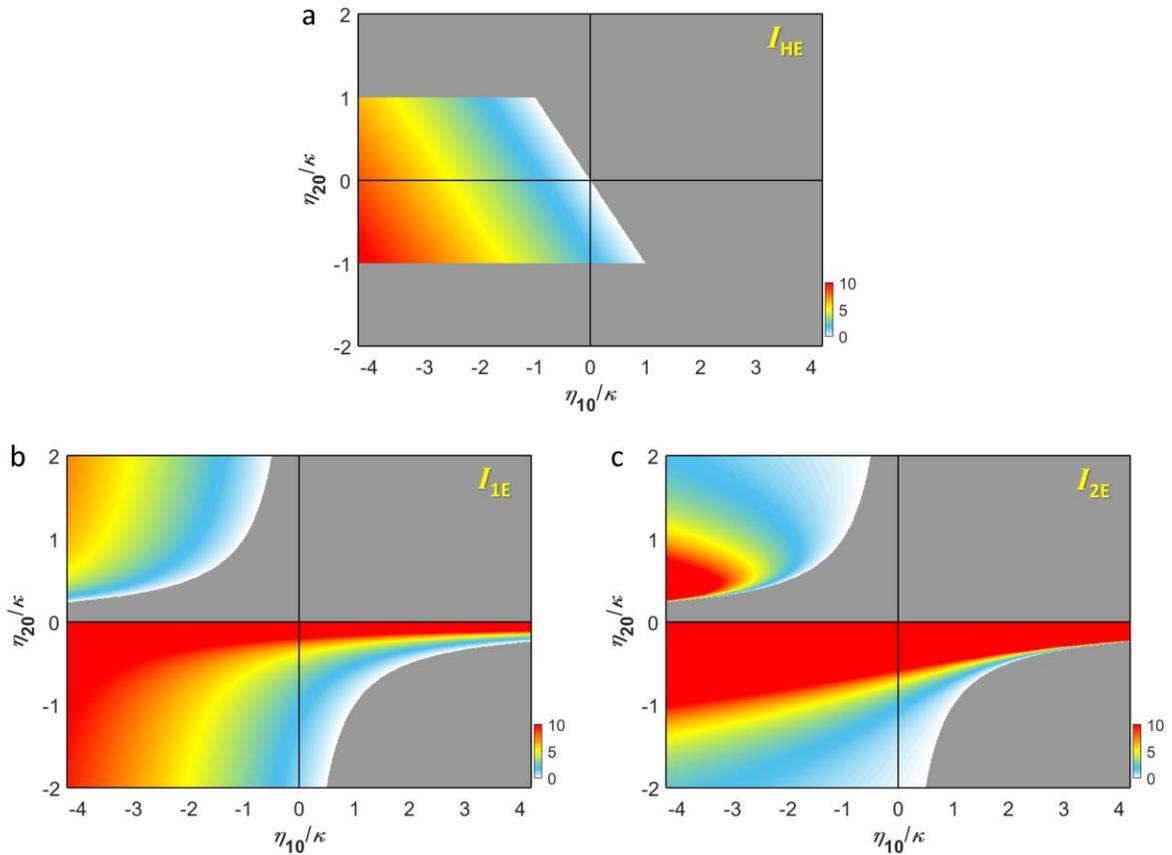

**Fig. S6.** Equilibria of TPE photonic molecules in the system parameter space $\eta_{10}$-$\eta_{20}$. (a) $I_{HE}$ for unbroken PT symmetry. (b) $I_{1E}$ and (c) $I_{2E}$ for broken PT symmetry. The gray areas denote the forbidden regions of nontrivial equilibria. $\eta_{11}/\kappa = +0.5$ for all cases.

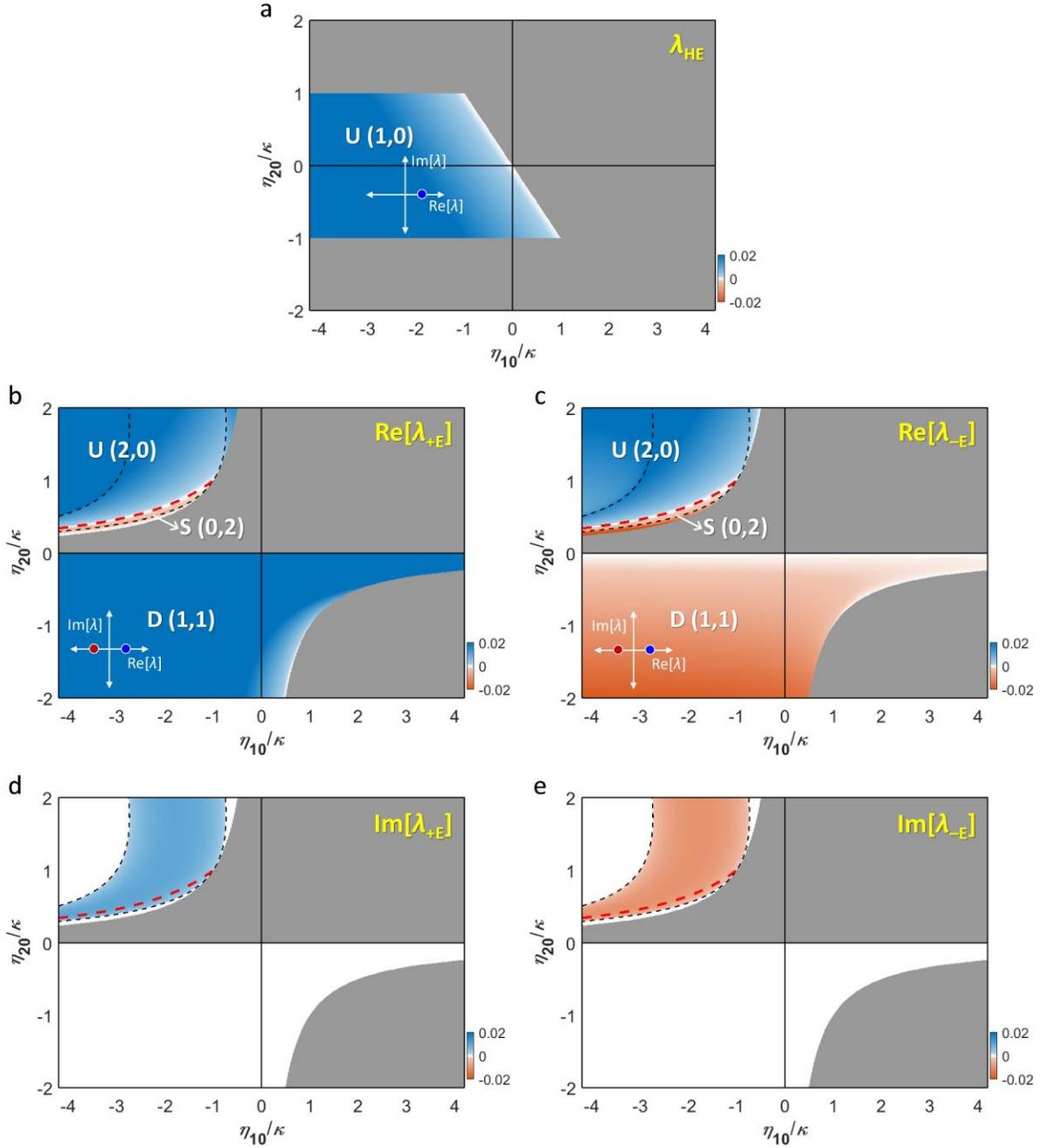

**Fig. S7.** Topological classification of TPE photonic molecules. (a) $\lambda_{HE}$ for unbroken PT symmetry. (b) Re[$\lambda_{+E}$], (c) Re[$\lambda_{-E}$], (d) Im[$\lambda_{+E}$], and (e) Im[$\lambda_{-E}$] for broken PT symmetry. The red lines denote the AH bifurcation, and the black dashed lines represent the boundaries between the node and focus phases. $\eta_{11}/\kappa = +0.5$ for all cases.

**Note S12. Topologically protected dynamics in the saddle and unstable phases**

The phase portraits in the saddle and unstable phases are presented in Fig. S8. While Fig. S8a shows the saddle phase dynamics obtained with the TPA resonator ($\eta_{11} < 0$), Fig. S8b and c represent the unstable dynamics obtained with TPE resonators ($\eta_{11} > 0$). Although both phases are unstable with respect to the nontrivial equilibrium, some of the initial states can converge to $(I_{1E}, I_{2E}) = (0,0)$, depending on the topological phase of the trivial equilibrium (0,0).

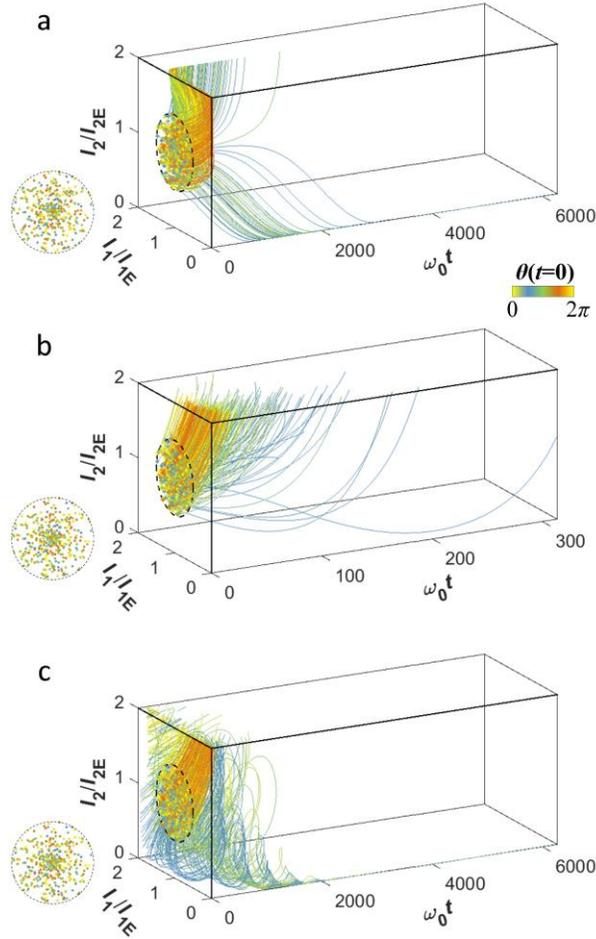

**Fig. S8.** Phase portraits of the saddle and unstable topological phases. (a-c) Trajectories of $(I_1, I_2)$. (a) D(1,1) phase with broken PT symmetry where $(\eta_{10}/\kappa, \eta_{20}/\kappa) = (-1.0, 0.5)$ and $\eta_{11}/\kappa = -0.5$. (b) U(0,2) phase with broken PT symmetry where $(\eta_{10}/\kappa, \eta_{20}/\kappa) = (-2.0, 1.5)$ and $\eta_{11}/\kappa = +0.5$. (c) U(0,1) phase with unbroken PT symmetry where $(\eta_{10}/\kappa, \eta_{20}/\kappa) = (-1.0, 0.5)$ and $\eta_{11}/\kappa = +0.5$. All other parameters are the same as those in Fig. 2 in the main text.

## Note S13. Temporal system perturbations

In Fig. 3 in the main text, temporal system perturbations are defined by the inclusion of random square pulses in $(\eta_{10}/\kappa, \eta_{20}/\kappa) = (1.0, -0.5)$ for the S(0,1) phase and $(\eta_{10}/\kappa, \eta_{20}/\kappa) = (2.0, -1.5)$ for the S(0,2) phase, forming $\eta_{10}(t)$ and $\eta_{20}(t)$ shown in Fig. S9. The bandwidths, intervals, and amplitudes of the square pulses have uniform random distributions.

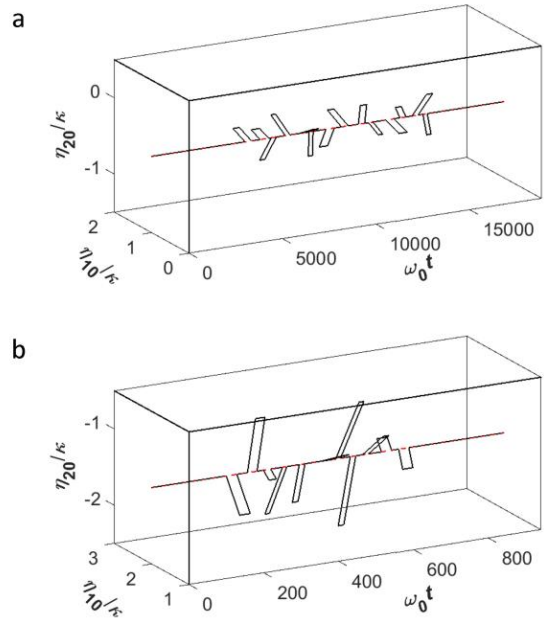

**Fig. S9.** Temporal variations in the system perturbations. (a,b) $\eta_{10}(t)$ and $\eta_{20}(t)$ for the (a) S(0,1) phase and (b) S(0,2) phase.

**Note S14. Noise in time-varying loss**

In the time domain analysis in Figs. 4 and 5 in the main text, sinusoidal and square pulses including noise components are applied to test noise-immune laser rectification and modulation, respectively. For the target signal $f_{target}(t)$ (square pulse in Fig. 4b and sinusoidal pulse in Fig. 5b in the main text), we set the modulation input $\eta_{20}(t)$ by adding a random perturbation $\delta_{20}(t)$ to $f_{target}(t)$ as $\eta_{20}(t) = f_{target}(t) + \delta_{20}(t)$, where:

$$\delta_{20}(t) = \int_W u[0,\delta]\cos(\omega t + u[-\pi,\pi])d\omega, \tag{S17}$$

$W = [\omega_L, \omega_H]$ is the spectral bandwidth of the noise component, and $u[p,q]$ is the uniform random function. The strength of the noise is then determined by the magnitude of $\delta$ (Fig. S10a for Figs. 4c and 5c, and Fig. S10b for Figs. 4d and 5d). In both examples, $\omega_L = 0.1\omega_0$ and $\omega_H = 0.2\omega_0$.

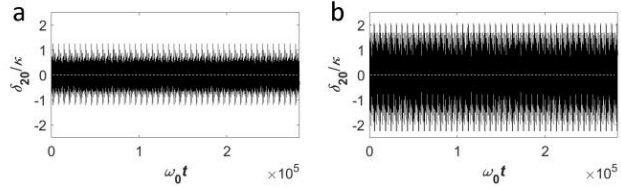

**Fig. S10.** Noise components $\delta_{20}(t)$ defined by the uniform random function and the magnitude $\delta$. (a) $\delta = 77.5 / \omega_0$ and (b) $\delta = 155 / \omega_0$.